\documentclass[twocolumn,showpacs]{revtex4}

\usepackage{graphics}

\begin{document}

\title{Influence of pump-field scattering to nonclassical-light generation in
a photonic band-gap nonlinear planar waveguide}

\author{Jan Pe\v{r}ina, Jr.}
\affiliation{Joint Laboratory of Optics of Palack\'{y} University
and Institute of Physics of Academy of Sciences of the Czech
Republic, 17. listopadu 50A, 772 07 Olomouc, Czech Republic}
\email{perina_j@sloup.upol.cz}
\author{Concita Sibilia}
\author{Daniela Tricca}
\author{Mario Bertolotti}
\affiliation{Dipartimento di Energetica, Universit\`{a} ``La
Sapienza'' di Roma, Via A. Scarpa 16, 00161 Roma, Italy}

\begin{abstract}
Optical parametric process occurring in a nonlinear planar
waveguide can serve as a source of light with nonclassical
properties. Properties of the generated fields are substantially
modified by scattering of the nonlinearly interacting fields in a
photonic band-gap structure inside the waveguide. A quantum model
of linear operator amplitude corrections to amplitude mean-values
provides conditions for an efficient squeezed-light generation as
well as generation of light with sub-Poissonian photon-number
statistics. Destructive influence of phase mismatch of the
nonlinear interaction can fully be compensated using a suitable
photonic-band gap structure inside the waveguide. Also an increase
of signal-to-noise ratio of an incident optical field can be
reached in the waveguide.
\end{abstract}


\maketitle

\section{Introduction}

Properties of linear photonic band-gap structures have been an
object of intensive investigations in the last several years
\cite{Bertolotti2001,Joannopoulos}. The most typical
characteristics of these structures are spatial localization of
optical modes in confined regions of a given structure and high
densities of these optical modes. Considering nonlinear materials,
energy of an optical field is mainly contained in these localized
modes and so a very strong and thus efficient nonlinear
interaction can occur. For example, second harmonic and
sub-harmonic generation in photonic band-gap structures has been
an object of investigation in \cite{Scalora1997,Dumeige2001}.
Phase-matching can be tailored in photonic band-gap structures so
that fulfilment of phase-matching conditions for a given nonlinear
process is reached and an efficient nonlinear process is
guaranteed this way. In some cases the overlap of nonlinearly
interacting optical fields and their mutual spatial phase
relations determine the strength of nonlinear process and
properties of light obtained in a nonlinear photonic band-gap
structure.

These properties may also be suitable for the generation of light
with nonclassical properties (squeezed light, light with
sub-Poissonian photon-number statistics), as has been suggested in
\cite{Sakoda2002}. Up to now, attention has been devoted to the
generation of nonclassical light in nonlinear photonic band-gap
waveguides. It has been shown that the process of second-harmonic
generation in a planar nonlinear waveguide with a corrugation on
the top can be used to control squeezing of the fundamental field
\cite{Tricca2004a}; the corrugation reproduces a photonic band-gap
structure. In \cite{Tricca2004a} periodicity of the grating was
selected to give rise to a longitudinal confinement of the pump
field, phase matching of the nonlinear process was achieved
introducing a spatial modulation of nonlinear susceptibility.
Conditions for an efficient squeezed-light generation as well as
generation of light with sub-Poissonian photon-number statistics
have been analyzed in \cite{PerinaJr2004} for a nonlinear
waveguide with optical parametric process; the photonic band-gap
structure was set to assure longitudinal confinement of the
down-converted fields.

In this contribution we extend the analysis given in
\cite{PerinaJr2004} to account also for the pump-field
longitudinal confinement. This confinement considerably changes
amplitude and phase relations along the waveguide thus providing
new possibilities for nonclassical-light generation. Optical
fields participating in the nonlinear interaction are described
using the generalized superposition of signal and noise.

A quantum derivation of the equations governing the evolution of
the interacting optical fields is given in Sec. 2. In Sec. 3,
conditions for squeezed-light generation are analyzed. Sec. 4 is
devoted to photon-number statistics of the generated fields.
Possibility to improve signal-to-noise ratio of an incident
optical field is discussed in Sec. 5. Sec. 6 contains conclusions.

\section{Quantum description of the nonlinearly interacting fields}

A quantum description of nonlinearly interacting optical modes
requires the construction of an appropriate momentum operator $
\hat{G}(z) $, which then determines Heisenberg equations of
motion:
\begin{equation}   
 \frac{d\hat{X}}{dz} = - \frac{i}{\hbar} \left[ \hat{G}, \hat{X}
  \right] ;
\label{1}
\end{equation}
$ \hat{X} $ stands for an arbitrary operator, $ \hbar $ is the
reduced Planck constant, and $ [\;,\;] $ means a commutator.

If a nonlinear interaction involves counter-propagating fields, we
cannot straightforwardly assign any momentum operator $ \hat{G} $
to the system of interacting optical fields. However, we can
proceed as follows \cite{PerinaJr2000}. We assume the nonlinear
interaction among all involved fields as if they co-propagate and
write the momentum operator $ \hat{G}(z) $ in the form:
\begin{eqnarray}  
 \hat{G}(z) = \sum_{a=s_F,i_F,p_F} \hbar ({\bf k}_a)_z \hat{a}^\dagger_a
 \hat{a}_a  + \sum_{a=s_B,i_B,p_B} \hbar ({\bf k}_a)_z \hat{a}^\dagger_a
 \hat{a}_a  \nonumber \\
 \mbox{} + \left[ \hbar K_{s} \exp(i\delta_l z)
  \hat{a}^\dagger_{s_F} \hat{a}_{s_B} + \hbar K_{i} \exp(i\delta_l z)
  \hat{a}^\dagger_{i_F} \hat{a}_{i_B} \right.
  \nonumber \\
  \mbox{} \left. + \hbar K_{p} \exp(2i\delta_l z)
  \hat{a}^\dagger_{p_F} \hat{a}_{p_B} + {\rm h.c.} \right] \nonumber \\
 \mbox{} - \left[ 2i\hbar K_{F} \hat{a}_{p_F}\hat{a}^\dagger_{s_F}
  \hat{a}^\dagger_{i_F} + 2i\hbar K_{B}
  \hat{a}_{p_B}\hat{a}^\dagger_{s_B} \hat{a}^\dagger_{i_B} +
  {\rm h.c.} \right] . \nonumber \\
\end{eqnarray}
Symbol $ \hat{a}_a $ ($ \hat{a}_a^\dagger $) denotes an
annihilation (creation) operator of mode $ a $; $ ({\bf k}_a)_z $
is the corresponding wave-vector of mode $ a $ along $ z- $axis.
We consider six nonlinearly interacting optical modes in the
investigated waveguide with optical parametric process;
forward-propagating signal mode (denoted as $ s_F $),
backward-propagating signal mode ($ s_B $), forward-propagating
idler mode ($ i_F $), backward-propagating idler mode ($ i_B $),
forward-propagating pump mode ($ p_F $), and finally
backward-propagating pump mode ($ p_B $). Constants $ K_s $, $ K_i
$, and $ K_p $ describe a linear exchange of energy between
forward- and backward-propagating signal, idler, and pump modes.
This exchange of energy originates in scattering of fields in a
photonic band-gap structure. Frequency $ \delta_l $ determines
periodicity of the corrugation on the top of the waveguide.
Constants $ K_F $ and $ K_B $ stand for nonlinear coupling
coefficients for forward- and backward-propagating fields. Values
of these parameters are determined using simple expressions after
a mode structure of the waveguide is found (see, e.g., in
\cite{Yeh1988,Pezzetta2001}). Scheme of the waveguide with the
considered interactions among the involved optical fields is shown
in Fig. \ref{1}.
\begin{figure}    
 \resizebox{0.7\hsize}{!}{\includegraphics{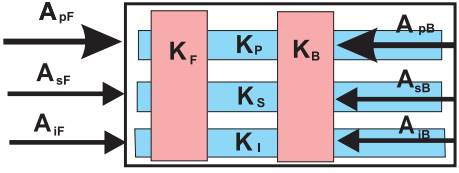}}
 \vspace{2mm}
 \caption{Scheme of the considered nonlinear planar photonic band-gap waveguide;
 $ A_{pF} $, $ A_{sF} $, and $ A_{iF} $ [$ A_{pB} $, $ A_{sB} $, and $ A_{iB}
 $] denote amplitudes of forward-propagating [backward-propagating]
 pump, signal, and idler fields; $ K_p $, $ K_s $, and $ K_i $ are
 linear coupling constants between pump, signal, and idler fields,
 and $ K_F $ ($ K_B $) stands for a nonlinear coupling constant
 among forward- (backward-) propagating fields.}
 \label{fig1}
\end{figure}

We now substitute creation operators ($ \hat{a}^\dagger $) of the
backward-propagating fields by newly introduced auxiliary
annihilation operators ($ \hat{b} $) and vice versa, i.e.
\begin{eqnarray}    
 \hat{a}^\dagger_{s_B} \leftarrow \hat{b}_{s_B} ,
 \hat{a}^\dagger_{i_B} \leftarrow \hat{b}_{i_B} ,
 \hat{a}^\dagger_{p_B} \leftarrow \hat{b}_{p_B} , \nonumber \\
 \hat{a}_{s_B} \leftarrow \hat{b}^\dagger_{s_B} ,
 \hat{a}_{i_B} \leftarrow \hat{b}^\dagger_{i_B} ,
 \hat{a}_{p_B} \leftarrow \hat{b}^\dagger_{p_B} .
\label{3}
\end{eqnarray}
Heisenberg equations in Eq. (\ref{1}) then have the form:
\begin{eqnarray}    
 \frac{d\hat{a}_{s_F}}{dz} &=& i ({\bf k}_{s_F})_z \hat{a}_{s_F}
  + i K_{s}\exp(i\delta_l z) \hat{b}^\dagger_{s_B}
  \nonumber \\
 & & \mbox{} + 2K_{F}\hat{a}_{p_F}\hat{a}^\dagger_{i_F},
  \nonumber \\
 \frac{d\hat{a}_{i_F}}{dz} &=& i ({\bf k}_{i_F})_z \hat{a}_{i_F}
  + i K_{i}\exp(i\delta_l z) \hat{b}^\dagger_{i_B}
  \nonumber \\
 & & \mbox{} + 2K_{F}\hat{a}_{p_F}\hat{a}^\dagger_{s_F},
  \nonumber \\
 \frac{d\hat{b}^\dagger_{s_B}}{dz} &=& -i ({\bf k}_{s_B})_z
  \hat{b}^\dagger_{s_B} - i K^*_{s}\exp(-i\delta_l z)
   \hat{a}_{s_F}
  \nonumber \\
 & & \mbox{} - 2K_{B}\hat{b}^\dagger_{p_B}\hat{b}_{i_B},
  \nonumber \\
 \frac{d\hat{b}^\dagger_{i_B}}{dz} &=& -i ({\bf k}_{i_B})_z
  \hat{b}^\dagger_{i_B} - i K^*_{i}\exp(-i\delta_l z)
   \hat{a}_{i_F}
  \nonumber \\
 & & \mbox{} - 2K_{B}\hat{b}^\dagger_{p_B}\hat{b}_{s_B},
  \nonumber \\
 \frac{d\hat{a}_{p_F}}{dz} &=& i ({\bf k}_{p_F})_z
  \hat{a}_{p_F} + i K_{p}\exp(2i\delta_l z) \hat{b}^\dagger_{p_B}
  \nonumber \\
 & & \mbox{}  - 2K^*_{F}\hat{a}_{s_F}\hat{a}_{i_F} ,
  \nonumber \\
 \frac{d\hat{b}^\dagger_{p_B}}{dz} &=& -i ({\bf k}_{p_B})_z
  \hat{b}^\dagger_{p_B} - i K_{p}^* \exp(-2i\delta_l z) \hat{a}_{p_F}
  \nonumber \\
 & & \mbox{}   + 2K^*_{B}\hat{b}^\dagger_{s_B}\hat{b}^\dagger_{i_B}.
\label{4}
\end{eqnarray}
To reach the final equations, we have to make the following steps:
1. Return to the original operators $ \hat{a}^\dagger $, $ \hat{a}
$ in Eqs. (\ref{4}) using the substitution in Eq. (\ref{3}). 2.
Transform Eqs. (\ref{4}) into the interaction picture ($
\hat{a}_a(z) = \hat{A}_a(z) \exp[i({\bf k}_a)_z z] $). 3. Write
operators $ \hat{A}_a(z) $ in the interaction picture as $
\hat{A}_a(z) = A_a(z) + \delta\hat{A}_a(z) $, where $ A_a(z) $ is
a classical amplitude mean-value and $ \delta\hat{A}(z) $ is a
small operator correction to this amplitude mean-value. This
procedure results in a system of nonlinear differential equations
for amplitude mean-values $ A_a $ and a system of linear operator
differential equations for small operator amplitude corrections $
\delta\hat{A}_a $.

The system of nonlinear differential equations for classical
amplitude mean-values $ A_a $ is written as follows:
\begin{eqnarray}   
\frac{dA_{s_F}}{dz} &=& iK_{s}\exp(-i\delta_{s}z) A_{s_B}
 \nonumber \\
 & & \mbox{} + 2K_{F} \exp(i\delta_{F}z) A_{p_F} A^*_{i_F},
\nonumber\\
\frac{dA_{i_F}}{dz} &=& iK_{i}\exp(-i\delta_{i}z) A_{i_B}
 \nonumber \\
 & & \mbox{} + 2K_{F} \exp(i\delta_{F}z) A_{p_F} A^*_{s_F},
\nonumber\\
\frac{dA_{s_B}}{dz} &=& -iK^*_{s}\exp(i\delta_{s}z) A_{s_F}
 \nonumber \\
 & & \mbox{} - 2K_{B} \exp(-i\delta_{B}z) A_{p_B} A^*_{i_B},
 \nonumber\\
\frac{dA_{i_B}}{dz} &=& -iK^*_{i}\exp(i\delta_{i}z) A_{i_F}
 \nonumber \\
 & & \mbox{} - 2K_{B} \exp(-i\delta_{B}z) A_{p_B} A^*_{s_B},
\nonumber\\
 \frac{dA_{p_F}}{dz} &=& iK_{p}\exp(-i\delta_{p}z) A_{p_B}
  \nonumber \\
  & & \mbox{} -2K^*_{F} \exp(-i\delta_{F}z)
   A_{s_F} A_{i_F},
\nonumber\\
 \frac{dA_{p_B}}{dz} &=& -iK^*_{p}\exp(i\delta_{p}z) A_{p_F}
  \nonumber \\
  & & \mbox{} + 2K^*_{B} \exp(i\delta_{B}z) A_{s_B}
    A_{i_B},
\label{5}
\end{eqnarray}
and
\begin{eqnarray}    
 \delta_{a} &=& |({\bf k}_{a_F})_z| + |({\bf k}_{a_B})_z| -
 \delta_l ,
  \hspace{0.5cm} a=s,i, \nonumber \\
 \delta_{p} &=& |({\bf k}_{p_F})_z| + |({\bf k}_{p_B})_z| -
 2\delta_l , \nonumber \\
 \delta_{b} &=& |({\bf k}_{p_b})_z| - |({\bf k}_{s_b})_z| -
 |({\bf k}_{i_b})_z| , \hspace{0.5cm} b=F,B .
\end{eqnarray}

Any solution of the system in Eqs. (\ref{5}) obeys the following
conservation law of energy (in quantum interpretation, ``the
overall number of virtual photons in the interaction'' is
conserved):
\begin{eqnarray}     
 \frac{d}{dz} \left( |A_{s_F}|^2 + |A_{i_F}|^2 + 2 |A_{p_F}|^2
  \right. \nonumber \\
   \left. - |A_{s_B}|^2 - |A_{i_B}|^2 -
  2 |A_{p_B}|^2 \right) = 0.
\end{eqnarray}

If the nonlinear terms in Eqs. (\ref{5}) are omitted, the solution
of Eqs. (\ref{5}) can be written as:
\begin{eqnarray}   
  A_{a_F}^{(0)} &=& \exp\left(-i\frac{\delta_a z}{2}\right) \left[B_a
   \cos(\Delta_a z) + \tilde{B_a} \sin(\Delta_a z) \right],
   \nonumber \\
  A_{a_B}^{(0)} &=& \exp\left(i\frac{\delta_a z}{2}\right) \nonumber \\
   & & \mbox{} \times  \left[B_a
   \left( -\frac{\delta_a}{2K_a}\cos(\Delta_a z) + i\frac{\Delta_a}{K_a}
   \sin(\Delta_a z) \right) \right.
   \nonumber \\
   & & \mbox{} \left. + \tilde{B_a} \left( -\frac{\delta_a}{2K_a}\sin(\Delta_a z)
   - i\frac{\Delta_a}{K_a} \cos(\Delta_a z) \right) \right],
   \nonumber \\
   & & a=s,i,p
   \label{8}
\end{eqnarray}
and
\begin{equation}  
  \Delta_a = \sqrt{\frac{\delta_a^2}{4} - |K_a|^2}; \hspace{1cm}
   a=s,i,p.
\end{equation}
In Eqs. (\ref{8}), constants $ B_{s} $, $ \tilde{B_{s}} $,
$B_{i}$, $ \tilde{B_{i}} $, $ B_{p} $, and $ \tilde{B_{p}} $ are
set according to boundary conditions at both sides of the
structure. The solution of the nonlinear set of equations written
in Eqs. (\ref{5}) is reached numerically using an iteration from
the solution written in Eqs.~(\ref{8}). A finite difference method
called BVP \cite{NumericalRecipes} has been found to be suitable
for this task.

The evolution of small operator amplitude corrections $
\delta\hat{A}_a $ is governed by the following equations:
\begin{eqnarray}     
\frac{d\delta \hat{A}_{s_F}}{dz} &=& {\cal K}_{s}
\delta\hat{A}_{s_B}
 + {\cal K}_{F}\left[ A_{p_F} \delta \hat{A}^\dagger_{i_F}
 + A^*_{i_F} \delta \hat{A}_{p_F} \right] ,
\nonumber\\
\frac{d\delta\hat{A}_{i_F}}{dz} &=& {\cal K}_{i}
\delta\hat{A}_{i_B}
  + {\cal K}_{F} \left[ A_{p_F} \delta\hat{A}^\dagger_{s_F}
 +  A^*_{s_F} \delta\hat{A}_{p_F}  \right] ,
\nonumber\\
\frac{d\delta\hat{A}_{s_B}}{dz} &=& {\cal K}^*_{s}
\delta\hat{A}_{s_F}
 - {\cal K}_{B} \left[ A_{p_B} \delta\hat{A}^*_{i_B}
  + A^*_{i_B} \delta\hat{A}_{p_B} \right],
 \nonumber\\
\frac{d\delta\hat{A}_{i_B}}{dz} &=& {\cal K}^*_{i}
\delta\hat{A}_{i_F}
 - {\cal K}_{B} \left[ A_{p_B}\delta\hat{A}^\dagger_{s_B}
  + A^*_{s_B}\delta\hat{A}_{p_B} \right],
\nonumber\\
 \frac{d\delta\hat{A}_{p_F}}{dz} &=& {\cal K}_{p}
 \delta\hat{A}_{p_B} -{\cal K}^*_{F}
  \left[ A_{s_F} \delta\hat{A}_{i_F} + A_{i_F}
   \delta\hat{A}_{s_F} \right],
\nonumber\\
 \frac{d\delta\hat{A}_{p_B}}{dz} &=& {\cal K}^*_{p}
  \delta\hat{A}_{p_F} + {\cal K}^*_{B} \left[ A_{s_B}
    \delta\hat{A}_{i_B} + A_{i_B} \delta\hat{A}_{s_B} \right] .
\label{10}
\end{eqnarray}
Functions $ {\cal K}_s $, $ {\cal K}_i $, $ {\cal K}_p $, $ {\cal
K}_F $, and $ {\cal K}_B $ introduced in Eqs. (\ref{10}) are
defined as:
\begin{eqnarray}   
 {\cal K}_a &=& iK_a \exp(-i\delta_a z), \hspace{1cm} a=s,i,p,
 \nonumber \\
 {\cal K}_F &=& 2K_F \exp(i\delta_F z),
 \nonumber \\
 {\cal K}_B &=& 2K_B \exp(-i\delta_B z).
\end{eqnarray}

The solution of the system of linear equations in Eqs. (\ref{10})
for operator amplitude corrections $ \delta\hat{A}_a $ can be
found numerically and written in the following matrix form:
\begin{equation}     
 \pmatrix{\delta \hat{\cal A}_{F,\rm out} \cr \delta \hat{\cal
 A}_{B,\rm in}}
 =  \pmatrix{ {\cal U}_{FF} & {\cal U}_{FB} \cr
 {\cal U}_{BF} & {\cal U}_{BB} }
 \pmatrix{\delta \hat{\cal A}_{F,\rm in} \cr
  \delta \hat{\cal A}_{B,\rm out}},
 \label{12}
\end{equation}
where
\begin{eqnarray}      
\delta \hat{\cal A}_{F,\rm in} = \pmatrix{\delta \hat{A}_{s_F}(0)
\cr \delta \hat{A}^\dagger_{s_F}(0) \cr \delta \hat{A}_{i_F}(0)
\cr \delta \hat{A}^\dagger_{i_F}(0) \cr \delta \hat{A}_{p_F}(0)
\cr \delta \hat{A}^\dagger_{p_F}(0)} , \hspace{0.5cm} \delta
\hat{\cal A}_{F,\rm out} = \pmatrix{ \delta \hat{A}_{s_F}(L) \cr
\delta \hat{A}^\dagger_{s_F}(L) \cr \delta \hat{A}_{i_F}(L) \cr
\delta \hat{A}^\dagger_{i_F}(L) \cr \delta \hat{A}_{p_F}(L) \cr
\delta \hat{A}^\dagger_{p_F}(L)} , \nonumber \\
\delta \hat{\cal A}_{B,\rm in} = \pmatrix{ \delta \hat{A}_{s_B}(L)
\cr \delta \hat{A}^\dagger_{s_B}(L) \cr \delta \hat{A}_{i_B}(L)
\cr \delta \hat{A}^\dagger_{i_B}(L) \cr \delta \hat{A}_{p_B}(L)
\cr \delta \hat{A}^\dagger_{p_B}(L) }, \hspace{0.5cm} \delta
\hat{\cal A}_{B,\rm out} = \pmatrix{ \delta \hat{A}_{s_B}(0) \cr
\delta \hat{A}^\dagger_{s_B}(0) \cr \delta \hat{A}_{i_B}(0) \cr
\delta \hat{A}^\dagger_{i_B}(0) \cr \delta \hat{A}_{p_B}(0) \cr
\delta \hat{A}^\dagger_{p_B}(0) } . \label{13}
\end{eqnarray}
Matrices $ {\cal U}_{FF} $, $ {\cal U}_{FB} $, $ {\cal U}_{BF} $,
and $ {\cal U}_{BB} $ characterize the solution of Eqs.
(\ref{10}).

Input-output relations among linear operator amplitude corrections
$ \delta\hat{A}_a $ can be found solving Eqs. (\ref{12}) with
respect to vectors $ \delta \hat{\cal A}_{F,\rm out} $ and $
\delta \hat{\cal A}_{B,\rm out} $:
\begin{eqnarray}     
 \pmatrix{\delta \hat{A}_{F,\rm out} \cr
 \delta \hat{A}_{B,\rm out}}  &=&
 \pmatrix{ {\cal U}_{FF}-{\cal U}_{FB} {\cal U}^{-1}_{BB}
 {\cal U}_{BF} & {\cal U}_{FB} {\cal U}^{-1}_{BB} \cr
 -{\cal U}^{-1}_{BB}{\cal U}_{BF} & {\cal U}^{-1}_{BB}}
 \nonumber \\
 & & \mbox{} \times
  \pmatrix{\delta \hat{\cal A}_{F,\rm in} \cr \delta \hat{\cal
  A}_{B,\rm in}}
  \label{14} \\
  &=&  {\cal U} \pmatrix{\delta \hat{\cal A}_{F,\rm in} \cr \delta \hat{\cal
  A}_{B,\rm in}}.
  \label{15}
\end{eqnarray}
Matrix $ {\cal U} $ defined in Eq. (\ref{15}) describes
input-output relations among the linear operator amplitude
corrections $ \delta\hat{A}_a $. The output linear operator
amplitude corrections contained in vectors $ \delta \hat{\cal
A}_{F,\rm out} $ and $ \delta \hat{\cal A}_{B,\rm out} $ obey
boson commutation relations provided that the input linear
operator amplitude corrections occurring in vectors $ \delta
\hat{\cal A}_{F,\rm in} $ and $ \delta \hat{\cal A}_{B,\rm in} $
obey boson commutation relations. It has been shown in
\cite{Luis1996} that this nontrivial property is fulfilled by any
system described by a quadratic hamiltonian.

The method of derivation of the operator equations in Eqs.
(\ref{10}) through the set of operator equations written in Eqs.
(\ref{4}) reveals that the following ``commutation relations''
among the small operator amplitude corrections $
\delta\hat{A}_a(L) $ are fulfilled:
\begin{eqnarray}    
 [\delta\hat{A}_i(L),\delta\hat{A}_k(L)] &=& 0,
  \nonumber \\
 {} [\delta\hat{A}_i(L),\delta\hat{A}^\dagger_k(L)] &=& \delta_{ik},
  \nonumber \\
 {} [\delta\hat{A}_i(L),\delta\hat{A}^\dagger_{\bar{k}}(L)] &=& 0,
  \nonumber \\
 {} [\delta\hat{A}_i(L),\delta\hat{A}_{\bar{k}}(L)] &=& 0,
  \nonumber \\
 {} [\delta\hat{A}_{\bar{i}}(L),\delta\hat{A}^\dagger_{\bar{k}}(L)] &=&
  -\delta_{\bar{i}\bar{k}},
  \nonumber \\
 {} [\delta\hat{A}_{\bar{i}}(L),\delta\hat{A}_{\bar{k}}(L)] &=& 0,
  \nonumber \\
   & & \hspace{-3cm} i,k=s_F,i_F,p_F, \hspace{1cm} \bar{i}, \bar{k} = s_B,i_B,p_B.
\label{16}
\end{eqnarray}
These relations have been found to be useful in controlling
precision of the numerical solution.

We describe the interacting fields in the framework of the
generalized superposition of signal and noise \cite{Perina1991}
(coherent states, squeezed states as well as noise can be
considered). Any state of a two-mode field is determined by values
of parameters $ B_j $, $ C_j $, $ D_{jk} $, and $ \bar{D}_{jk} $
\cite{PerinaJr2000}:
\begin{eqnarray}   
B_j &=& \langle \Delta\hat{A}^\dagger_j \Delta\hat{A}_j \rangle ,
  \nonumber \\
C_j &=& \langle (\Delta\hat{A}_j)^2 \rangle ,
  \nonumber \\
D_{jk} &=& \langle \Delta\hat{A}_j \Delta\hat{A}_k \rangle ,
 \hspace{1cm} j \neq k,  \nonumber \\
\bar{D}_{jk} &=& -\langle \Delta\hat{A}^\dagger_j \Delta\hat{A}_k
\rangle , \hspace{1cm} j \neq k;
\label{17}
\end{eqnarray}
$ \Delta\hat{A}_j = \hat{A}_j - \langle\hat{A}_j\rangle $. Symbol
$ \langle \;\; \rangle $ stands for a quantum statistical mean
value. Coefficients $ B_j $, $ C_j $, $ D_{jk} $, and $
\bar{D}_{jk} $ can then be determined using matrix $ \cal U $
introduced in Eq. (\ref{15}) and incident values of $ B_{j,\rm
in,{\cal A}} $ and $ C_{j,\rm in,{\cal A}} $ related to
anti-normal ordering of field operators (for details, see
\cite{PerinaJr2000}):
\begin{eqnarray}   
 B_{j,\rm in,{\cal A}} &=& \cosh^2(r_{j}) + n_{ch,j} ,
  \nonumber \\
 C_{j,\rm in,{\cal A}} &=& \frac{1}{2} \exp (i\vartheta_j)
  \sinh (2r_j) .
\end{eqnarray}
Symbol $ r_j $ denotes a squeeze parameter of the incident $ j
$-th mode, $ \vartheta_j $ means a squeeze phase, and $ n_{ch,j} $
stands for a mean number of incident chaotic photons. Coefficients
$ D_{jk,\rm in,{\cal A}} $ and $ \bar{D}_{jk,\rm in,{\cal A}} $
for an incident field are set to zero because the incident fields
are assumed to be statistically independent.

The expressions for coefficients $ B_j $, $ C_j $, $ D_{jk} $, and
$ \bar{D}_{jk} $ can be written in terms of matrix elements of $
\cal U $ as follows \cite{PerinaJr2000}:
\begin{eqnarray}   
 B_j &=& \sum_{k=1}^{6} \Bigl[ \left( {\cal U}^*_{2j-1,2k-1} {\cal U}_{2j-1,2k}
   C^*_{k,\rm in,{\cal A}} + {\rm c.c.} \right)
   \nonumber \\
   & & \mbox{} \hspace{-5mm} + |{\cal U}_{2j-1,2k-1}|^2
  \left( B_{k,\rm in,{\cal A}} -1 \right) +
  |{\cal U}_{2j-1,2k}|^2 B_{k,\rm in,{\cal A}}
    \Bigr],
   \nonumber \\
 C_j &=& \sum_{k=1}^{6} \Bigl[ {\cal U}^2_{2j-1,2k-1} C_{k,\rm in,{\cal A}}
  + {\cal U}^2_{2j-1,2k} C^*_{k,\rm in,{\cal A}}
   \nonumber \\
   & & \mbox{} + {\cal U}_{2j-1,2k-1}{\cal U}_{2j-1,2k}
   \left(2B_{k,\rm in,{\cal A}} -1 \right) \Bigr] ,
   \nonumber \\
 D_{jk} &=& \sum_{l=1}^{6} \Bigl[ {\cal U}_{2j-1,2l-1}{\cal U}_{2k-1,2l-1}
   C_{l,\rm in,{\cal A}} \nonumber \\
   & & \mbox{} + {\cal U}_{2j-1,2l}{\cal U}_{2k-1,2l}
   C^*_{l,\rm in,{\cal A}} \nonumber \\
   & & \mbox{} + {\cal U}_{2j-1,2l-1} {\cal U}_{2k-1,2l}
   B_{l,\rm in,{\cal A}}  \nonumber \\
   & & \mbox{} +
   {\cal U}_{2j-1,2l} {\cal U}_{2k-1,2l-1} \left( B_{l,\rm in,{\cal A}}
   -1 \right) \Bigr] , \;\; j\neq k,  \nonumber \\
\bar{D}_{jk} &=& \sum_{l=1}^{6} \Bigl[- {\cal U}^*_{2j-1,2l}
  {\cal U}_{2k-1,2l-1}  C_{l,\rm in,{\cal A}}
   \nonumber \\
   & & \mbox{}  -
   {\cal U}^*_{2j-1,2l-1} {\cal U}_{2k-1,2l} C^*_{l,\rm in,{\cal A}}
    \nonumber \\
   & & \mbox{} - {\cal U}^*_{2j-1,2l-1}
   {\cal U}_{2k-1,2l-1} \left( B_{l,\rm in,{\cal A}} - 1 \right)
   \nonumber \\
   & & \mbox{} - {\cal U}^*_{2j-1,2l}{\cal U}_{2k-1,2l}
   B_{l,\rm in,{\cal A}}
   \Bigr] , \;\; j\neq k;
 \label{19}
\end{eqnarray}
$ {\rm c.c.} $ stands for complex conjugated terms.

\section{Squeezed-light generation}

The level of noise present in quadrature components $ \hat{q}_j $
[$ \hat{q}_j = \hat{A}_j + \hat{A}^\dagger_j $, $ \hat{A}_j $
stands for an electric-field-amplitude operator of mode $ j $] and
$ \hat{p}_j $ [$ \hat{p}_j = -i(\hat{A}_j - \hat{A}^\dagger_j) $]
appropriate for mode $ j $ can be lower than the level of noise
characterizing the vacuum field. Then we speak about squeezed
light. In general, the maximum amount of available squeezing is
reached under some chosen value of a local-oscillator phase in the
homodyne-measurement scheme and the corresponding amount of
squeezing is given in theory by principal squeeze variance $
\lambda_j $ \cite{Luks1988}.

It is useful to combine some optical fields on a beam-splitter and
to study properties of the output fields. Such fields can have a
nonclassical character under certain conditions. For example,
signal and idler fields generated in optical parametric process
have this property, because one signal photon and one idler photon
are created together in one elementary quantum event of the
nonlinear process. We use the notation compound mode for this case
and define the appropriate operators for quadrature components
combining $ j $-th and $ k $-th  modes; $ \hat{q}_{jk} = \hat{q}_j
+ \hat{q}_k $ and $ \hat{p}_{jk} = \hat{p}_j + \hat{p}_k $.

Using the parameters characterizing a state and determined in Eqs.
(\ref{19}), we obtain for single-mode principal squeeze variance $
\lambda_j $ and compound-mode principal squeeze variance $
\lambda_{ij} $ the following expressions \cite{PerinaJr2000}:
\begin{eqnarray}     
\lambda_j &=& 1+ 2[B_j-|C_j|], \\
\lambda_{jk} &=& 2\left[1+ B_j+B_k -
  2{\rm Re}(\bar{D}_{jk}) \right. \nonumber \\
  & & \mbox{} \left. - | C_j+C_k+2D_{jk}| \right] .
\end{eqnarray}
Values of principal squeeze variance $ \lambda_j $ less than one
indicate squeezing in a single-mode case. Squeezed light is
generated in a compound-mode (two-mode) case if values of
principal squeeze variance $ \lambda_{jk} $ are less than two.

Similarly as in \cite{PerinaJr2004}, assuming $ K_s L $, $ K_i L
$, $ K_p L $, $ K_F A L $ ($ A $ being a typical field amplitude),
and $ K_B AL $ being small, analytical expressions for principal
squeeze variances can be found solving equations in Eqs.
(\ref{10}) iteratively. The obtained expressions for principal
squeeze variances for single-mode and compound-mode cases are the
same as those given in Eqs. (23) and (25) of \cite{PerinaJr2004},
where the system with $ K_p = 0 $ is analyzed. Thus, no
information about the influence of linear pump scattering on
squeezed-light generation can be obtained.

In the following discussion a strong incident forward-propagating
pump field and also nonzero incident forward-propagating signal
and idler fields are considered. Squeezed light cannot be
generated in single-mode cases. However, in general, compound
modes ($ s_F,i_F $), ($ s_B,i_B $), and ($ s_F,i_B $) generate
squeezed light under certain conditions. We note, that properties
of the optical fields that follow from the symmetry between the
signal and idler fields are not mentioned explicitly. For example,
following the symmetry, also compound mode ($ i_F,s_B $) can be
squeezed. It is interesting to note, that the analyzed waveguide
behaves qualitatively similarly as a waveguide with two separated
parts with optical parametric processes and with fields in
different parts interacting linearly through evanescent waves.
This waveguide with co-propagating fields has been analyzed in
\cite{Herec1999,Mista1999}.

\begin{figure}    
 \resizebox{0.7\hsize}{!}{\includegraphics{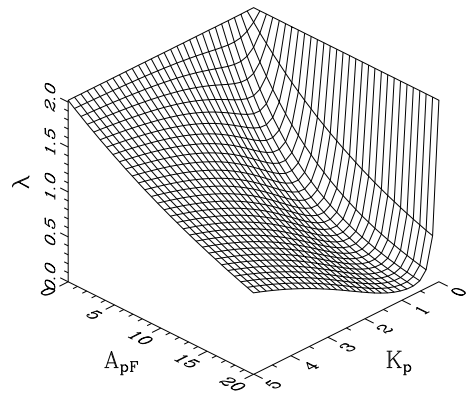}}

 \vspace{2mm}
 \caption{Principal squeeze variance $ \lambda $
 of mode ($ s_B,i_B $) as a function of forward-propagating pump
 amplitude $ A_{p_F} $ and linear coupling constant $ K_p $ between
 the pump modes;
 $ K_F = K_B = 5 \times 10^{-2} $,
 $ K_s = K_i = 0 $, $ \delta_s = \delta_i = 0 $, $ \delta_p =0 $,
 $ \delta_F = \delta_B = 0 $, $ A_{s_F} = A_{i_F} = 0.1 $,
 $ A_{p_B} = A_{s_B} = A_{i_B} =0 $;
 incident coherent states are assumed.
 The following units are used for the considered physical quantities:
 $ [K_F]=[K_B] = 10^{-6} {\rm mm}^{-1}{\rm m V}^{-1}, [K_s]=[K_i]=[K_p]=
 {\rm mm}^{-1}, [L] = {\rm mm}, [\delta_s]=[\delta_i]=[\delta_p]=
 [\delta_F]=[\delta_B]={\rm mm}^{-1},
 [A_a] = 10^{6} {\rm Vm}^{-1} $.}
 \label{fig2}
\end{figure}

We first consider linear scattering only between the pump modes ($
K_p \neq 0 $). If there is phase matching of all interactions, the
greater the value of $ K_p $ the greater the values of principal
squeeze variance $ \lambda $ of mode ($ s_F,i_F $). This means
that phases of the nonlinearly interacting optical fields along
the structure are modified by nonzero values of $ K_p $ in such a
way that the amount of generated squeezing decreases. On the other
hand, linear coupling given by nonzero values of $ K_p $ transfers
energy into mode $ p_B $ and so squeezing can be observed also in
mode ($ s_B,i_B $). However, greater values of $ K_p $ suppress
the amount of squeezing in this mode, as is shown in Fig.
\ref{fig2}.

In general, the greater the pump-field amplitude $ A_{p_F} $ the
lower the value of principal squeeze variance $ \lambda $.
Nonlinear coupling constants $ K_F $ and $ K_B $ behave in the
same way. Also the dependence of principal squeeze variances $
\lambda $ on length $ L $ of the waveguide is similar, because
``the overall amount of nonlinear interaction'' is proportional to
$ L $.

\begin{figure}    
 \resizebox{0.7\hsize}{!}{\includegraphics{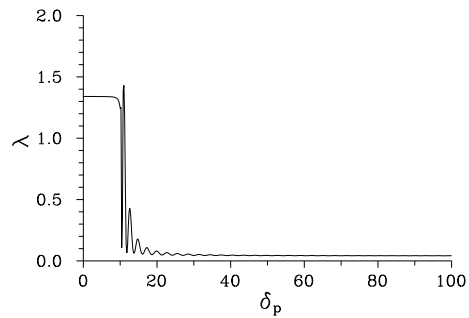}}

 \vspace{2mm}
 \caption{Principal squeeze variance $ \lambda $
 of mode ($ s_F,i_F $) as a function of linear phase mismatch $ \delta_p $ between
 the pump fields; $ L=2 $, $ K_p = 5 $, $ A_{p_F} = 10 $, and values of the
 other parameters are the same as in Fig. 2.}
 \label{fig3}
\end{figure}

If the waveguide is characterized by a greater value of $ K_p $
then the value of $ \lambda $ depends strongly on linear phase
mismatch $ \delta_p $ of the forward- and backward-propagating
pump fields. A dramatic decrease of principal squeeze variance $
\lambda $ in mode ($ s_F,i_F $) for $ \delta_p > 2|K_p| $ is shown
in Fig. \ref{fig3}. As the approximate solution in Eqs. (\ref{8})
for amplitude mean-values suggests, this region of parameters with
low values of $ \lambda $ can be characterized by an oscillating
behavior of the amplitudes (as functions of the position in the
waveguide). A similar decrease of $ \lambda $ occurs also in mode
($ s_B,i_B $) as $ \delta_p $ exceeds $ 2|K_p| $, but for greater
values of $ \delta_p $ an increase of values of $ \lambda $
follows because transfer of energy to the backward-propagating
pump field decreases.

\begin{figure}    
 \resizebox{0.8\hsize}{!}{\includegraphics{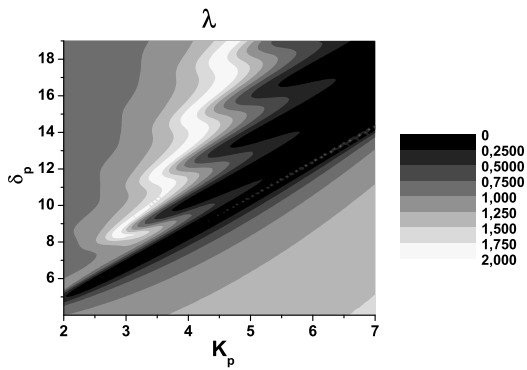}}

 \vspace{2mm}
 \caption{Topological graph of principal squeeze variance $ \lambda $
 in mode ($ s_F,i_F $) as it depends on linear coupling constant $ K_p $
 and linear phase mismatch $ \delta_p $ between
 the pump fields;
 $ L = 2 $, $ A_{p_F} = 10 $, $ \delta_F = \delta_B = 5 $,
 and values of the other parameters are the same as in Fig. 2.}
 \label{fig4}
\end{figure}

If the nonlinear interaction is not phase-matched ($ \delta_{nl} =
\delta_F =\delta_B \neq 0 $), then the linear scattering of the
pump field (described by $ K_p $ and $ \delta_p $) can compensate
for nonzero values of nonlinear phase-mismatch $ \delta_{nl} $ and
enable better values of principal squeeze variances $ \lambda $.
If, for example, we consider $ \delta_{nl}=5 $ mm${}^{-1} $ and
values of the other parameters written in caption to Fig.
\ref{fig4}, principal squeeze variance $ \lambda $ of mode ($
s_F,i_F $) is greater than 0.8 assuming the waveguide without a
photonic band-gap structure. Having a photonic band-gap structure
with suitable values of parameters inside the waveguide, principal
squeeze variance $ \lambda $ can even reach the value 0.04
appropriate also for a phase-matched nonlinear interaction in the
waveguide without a photonic band-gap structure. Suitable values
of $ K_p $ and $ \delta_p $ lie typically around the curve $
\delta_p = 2|K_p| $, as can be seen in Fig. \ref{fig4}. Also
greater values of $ K_p $ are needed to have lower values of $
\lambda $.
The possibility to compensate for a nonlinear phase-mismatch using
linear coupling between modes has been discussed in
\cite{PoDong2004} from the point of view of efficiency of energy
conversion.

Nonlinear phase mismatch $ \delta_{nl} $ can be also compensated
using nonzero incident amplitudes of the backward-propagating pump
mode. We have found the optimum value of the phase difference
between the forward-propagating pump amplitude and
backward-propagating pump amplitude to be $ 3\pi/2 $ [$ \pi/2 $]
for mode ($ s_F,i_F $) [($ s_B,i_B $)] with respect to
squeezed-light generation.

Now we assume linear scattering of pump, signal, and idler fields
($ K_p \neq 0 $, $ K_s \neq 0 $, $ K_i \neq 0 $). The analysis of
squeezed-light generation is difficult under these conditions
because every linear coupling constant introduces some phase
changes along the waveguide that modify the nonlinear interaction.
In general, squeezing can be observed in modes ($ s_F,i_F $), ($
s_B,i_B $), and ($ s_F,i_B $) under suitably chosen values of
parameters. The occurrence of squeezing in mode ($ s_F,i_B $)
originates in linear scattering of the down-converted fields.

Nonzero values of $ K_s $ and $ K_i $ support squeezed-light
generation in mode ($ s_B,i_B $), because they transfer energy
from modes $ s_F $ and $ i_F $ into modes $ s_B $ and $ i_B $ and
so the nonlinear interaction among the backward-propagating fields
can have also a strong stimulated part. A nonzero value of $ K_i $
is also indispensable for observing squeezing in mode ($ s_F,i_B
$), because ``an already generated squeezed light in the nonlinear
interaction among the forward-propagating fields has to be
transferred into mode $ i_B $''.

\begin{figure}    
 \resizebox{0.7\hsize}{!}{\includegraphics{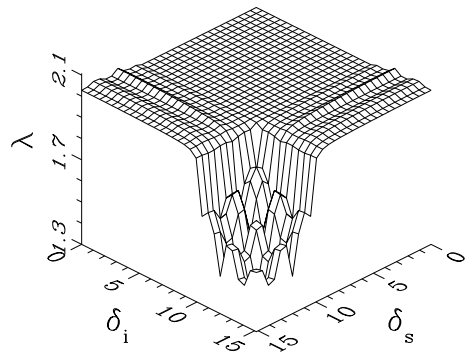}}

 \vspace{2mm}
 \caption{Principal squeeze variance $ \lambda $
 of mode ($ s_F,i_F $) as a function of linear phase mismatches $ \delta_s $
 and $ \delta_i $ between the signal and idler fields, respectively;
 $ L=2 $, $ K_p = 5 $, $ K_s = K_i = 5 $,  $ A_{p_F} = 10 $, and values of the
 other parameters are the same as in Fig. 2.}
 \label{fig5}
\end{figure}

Being in non-oscillating regime of behavior of amplitude
mean-values ($ \delta_p < 2|K_p| $, $ \delta_s < 2|K_s| $, $
\delta_i < 2|K_i| $) lower values of $ \lambda $ can be reached in
mode ($ s_B,i_B $) in comparison with those reached in mode ($
s_F,i_F $) probably because modes $ s_B $ and $ i_B $ begin the
nonlinear interaction in the vacuum states and so phases of their
amplitudes can be suitably set in the interaction. In general, the
regime with the oscillating behavior of amplitude mean-values ($
\delta_p
> 2|K_p| $, $ \delta_s > 2|K_s| $, $ \delta_i > 2|K_i| $) is
better for squeezed-light generation and low values of principal
squeeze variances $ \lambda $ can be reached in this regime.
Squeezing can be observed under a wide range of values of
parameters in non-oscillating regime in modes ($ s_F,i_F $), ($
s_B,i_B $), and ($ s_F,i_B $). We note, that greater values of,
e.g., $ \delta_i $ effectively decrease the value of $ K_i $ which
acts against squeezed-light generation in mode ($ s_F,i_B $). A
difference in the ability to generate squeezed-light in
oscillating and non-oscillating regimes is clearly visible in Fig.
\ref{fig5}, where principal squeeze variance $ \lambda $ of mode
($ s_F,i_B $) is plotted as a function of linear phase mismatches
$ \delta_s $ and $ \delta_i $.

As can be shown directly by a suitable substitution into Eqs.
(\ref{5}) and (\ref{10}), their solutions depend only on the
overall phase $ \psi = -\arg(K_p) + \arg(K_s) + \arg(K_i) $ of the
linear coupling constants. In our investigations, principal
squeeze variances $ \lambda $ in modes ($ s_F,i_F $) and ($
s_B,i_B $) had minimum values for $ \psi = \pi/2 $ and $ \lambda $
in mode ($ s_F,i_B $) reached its lowest value for $ \psi = 3\pi/2
$. Also the dependence of principal squeeze variances $ \lambda $
on phases of the incident forward-propagating signal and idler
fields have not been observed.

At the end of the discussion of squeezing we mention the case in
which, e.g., the incident forward-propagating signal field is
stronger than the incident forward-propagating pump field. We then
have squeezing also in compound modes that combine one pump field
with one down-converted field under suitably chosen values of
parameters. In general, values of principal squeeze variances $
\lambda $ are lower in these compound modes that contain a
down-converted field with great values of amplitudes inside the
structure.

\section{Photon-number statistics}

When a field is detected by a classical detector statistical
properties of the incident field are suitably described by
normally-ordered moments of integrated intensity $ \hat{W} $ ($
\hat{W} = \hat{A}^\dagger \hat{A} $):
\begin{equation}     
 \langle W^k \rangle_{\cal N} = \langle \hat{A}^{\dagger k}
  \hat{A}^k \rangle , \hspace{10mm} k=2,3,\ldots,
\end{equation}
where $ \hat{A} $ denotes an electric-field-amplitude operator of
the incident optical field.

Type of a statistical distribution of photoelectrons emitted
inside the detector can be determined using Fano factor $ F_n $.
Fano factor $ F_n $  can be expressed in terms of normally-ordered
moments of integrated intensity $ \hat{W} $ as follows:
\begin{equation}     
 F_n = \frac{ \langle (\Delta n)^2 \rangle }{ \langle n \rangle}
 = 1 + \frac{ \langle (\Delta W)^2 \rangle_{\cal N} }{
  \langle W \rangle_{\cal N} } .
 \label{23}
\end{equation}
Symbol $ n $ denotes the number of photoelectrons, $ \Delta n = n
- \langle n \rangle $,  and $ \Delta W = W - \langle W
\rangle_{\cal N} $. Intensity operator $ \hat{W}_{ij} $ of a
compound mode ($ i,j $) is then determined along the relation $
\hat{W}_{ij} = \hat{W}_i + \hat{W}_j $, where $ \hat{W}_i $ ($
\hat{W}_j $) stands for the intensity operator of mode $ i $ ($ j
$). Classical fields obey the inequality $ F_n \ge 1 $. On the
other hand values of Fano factor $ F_n $ smaller than one can be
reached considering nonclassical fields (sub-Poissonian light).
The condition $ F_n \le 1 $ means that fluctuations in the number
of photoelectrons are suppressed below the classical limit that is
given by the Poissonian photon-number statistics of a laser
radiation.

Assuming that the outgoing fields can be described in the
framework of the generalized superposition of signal and noise,
moments of integrated intensity $ \hat{W} $ can be written as:
\begin{eqnarray}    
 \langle W_j \rangle_{\cal N} &=& B_j + |\xi_j|^2 ,
  \nonumber \\
 \langle (\Delta W_j)^2 \rangle_{\cal N} &=& B_j^2 + |C_j|^2 +
  2B_j|\xi_j|^2 + \left(C_j\xi_j^{*2} + {\rm c.c.} \right) ,
   \nonumber \\
 \langle \Delta W_j \Delta W_k \rangle_{\cal N} &=& |D_{jk}|^2 +
   |\bar{D}_{jk}|^2 \nonumber \\
    & & \mbox{} + \left( D_{jk}\xi_j^*\xi_k^* - \bar{D}_{jk}\xi_j\xi_k^*
   + {\rm c.c.} \right) .
\label{24}
\end{eqnarray}
Coefficients $ B_j $, $ C_j $, $ D_{jk} $, and $ \bar{D}_{jk} $
are given in Eqs. (\ref{19}). Symbol $ \xi_j $ in Eqs. (\ref{24})
denotes a coherent amplitude of the $ j $th outgoing field. We
note that photon-number distribution as well as moments of
integrated intensity can be determined in general using an
expansion into Laguerre polynomials (see, e.g.,
\cite{Perina1991,Perinova1981}).

Nonclassical character of photon-number statistics occurs mostly
at single-photon level. For this reason, we assume a regime in
which the waveguide is pumped by a strong forward-propagating pump
field and classical strong amplitude mean-values $ A_{s_F} $, $
A_{i_F} $, $ A_{s_B} $, and $ A_{i_B} $ are zero. Operator
amplitudes $ \hat{A} $ of the signal and idler fields are then
given just by their linear operator amplitude corrections $ \delta
\hat{A} $. We also assume that the incident fields described only
by linear operator amplitude corrections are coherent and denote
their amplitudes by $ \xi $.

Assuming $ K_s L $, $ K_i L $, $ K_p L $, $ K_F A L $ ($ A $ being
a typical field amplitude), and $ K_B AL $ being small,
expressions for first and second moments of integrated intensity $
\hat{W} $ can be found as it was done in \cite{PerinaJr2004}. This
approximation gives the same expressions as those published in
Eqs. (30) and (31) in \cite{PerinaJr2004} for the case $ K_p = 0
$; i.e. no conclusion about the influence of $ K_p $ can be
deduced.

\begin{figure}    
 \resizebox{0.7\hsize}{!}{\includegraphics{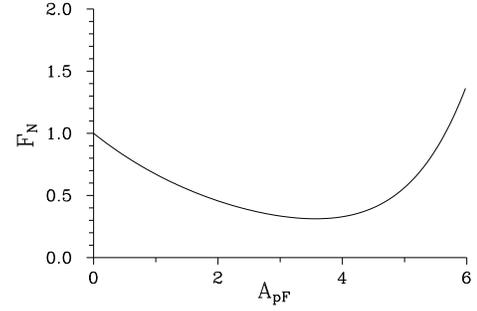}}

 \vspace{2mm}
 \caption{Fano factor $ F_n $
 of mode ($ s_F,i_F $) as a function of
 forward-propagating pump amplitude $ A_{p_F} $;
 input linear operator amplitude corrections $ \delta \hat{A} $
 are assumed to be in coherent states
 with amplitudes $ \xi $, amplitudes $ \xi $ are expressed in
 units of 10 Vm$ {}^{-1} $ and so mean values of intensities
 are directly equal to mean photon numbers;
  $ L=2 $, $ K_F = K_B = 5 \times 10^{-2} $,
  $ K_p = K_s = K_i = 0 $, $ \delta_p = \delta_s = \delta_i = 0 $,
  $ \delta_F = \delta_B = 0 $, $ A_{p_F} = 10 $,
  $ A_{s_F} = A_{i_F} = 0 $, $ A_{p_B} = A_{s_B} = A_{i_B} =0 $;
  $ \xi_{s_F} = -10 $, $ \xi_{i_F} = 10 $
  $ \xi_{p_F} = \xi_{s_B} = \xi_{i_B} = \xi_{p_B} = 0 $.}
 \label{fig6}
\end{figure}

\begin{figure}    
 \resizebox{0.7\hsize}{!}{\includegraphics{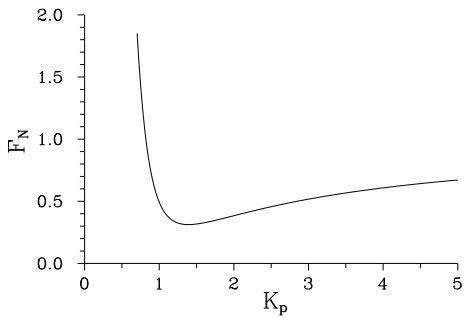}}

 \vspace{2mm}
 \caption{Fano factor $ F_n $
 of mode ($ s_F,i_F $) as a function
 linear coupling constant $ K_p $ between the pump fields;
 $ A_{p_F} = 10 $ and
 values of the other parameters are the same as in Fig. 6.}
 \label{fig7}
\end{figure}

We first consider a phase-matched nonlinear waveguide with no
photonic band-gap structure ($ K_s = K_i = K_p = 0 $).
Sub-Poissonian light in mode ($ s_F,i_F $) can occur for a
sufficiently strong pumping. However, as Fig. \ref{fig6} shows, if
the value of pump amplitude $ A_{p_F} $ is too great,
sub-Poissonian character of the generated light is lost.
Suitable values of pump amplitude $ A_{p_F} $ for
sub-Poissonian-light generation depend on length $ L $ of the
waveguide. The longer the waveguide, the smaller the suitable
values of pump amplitude $ A_{p_F} $. A photonic band-gap
structure with $ K_p \neq 0 $ (also $ K_s = K_i = 0 $ is assumed)
inside the waveguide in this case leads to a redistribution of
energy in the pump modes in the way that enables again
sub-Poissonian-light generation in mode ($ s_F,i_F $). For a given
value of length $ L $ and a given sufficiently great value of $
A_{p_F} $ there is an optimum value of $ K_p $ for which the value
of Fano factor $ F_n $ reaches a minimum value that is obtained
also in the waveguide without a photonic band-gap structure ($ F_n
\approx 0.3 $, see Fig. \ref{fig7}). If the value of pump
amplitude $ A_{p_F} $ is small, the greater the $ K_p $ the
greater the Fano factor $ F_n $; i.e. a photonic band-gap
structure does not support sub-Poissonian-light generation in this
case.
Nonzero values of linear phase mismatch $ \delta_p $ in this
otherwise phase-matched interaction result in greater values of
Fano factor $ F_n $.

\begin{figure}    
 \resizebox{0.7\hsize}{!}{\includegraphics{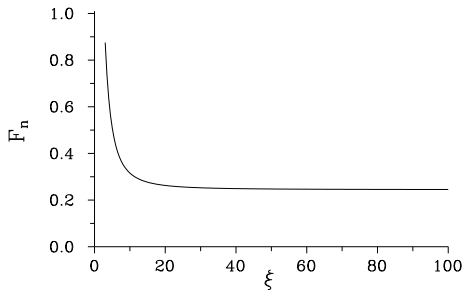}}

 \vspace{2mm}
 \caption{Fano factor $ F_n $
 of mode ($ s_F,i_F $) as a function of amplitude $ \xi $:
 $ \xi_{s_F} = - \xi $, $ \xi_{i_F} = \xi $:
 $ K_p = 1.4 $, $ A_{p_F} = 10 $,
 and values of the other parameters are the same as
 in Fig. 6.}
 \label{fig8}
\end{figure}

In order to obtain sub-Poissonian light in mode ($ s_F,i_F $), the
nonlinear interaction has to be stimulated, i.e. the incident
small amplitudes $ \xi_{s_F} $ and $ \xi_{i_F} $ have to be
nonzero. Moreover, values of amplitudes $ \xi_{s_F} $ and $
\xi_{i_F} $ have to be nearly the same and their phases have to
fulfill the condition $ \arg(\xi_{s_F}) + \arg(\xi_{i_F}) \approx
\pi $. Under these conditions Fano factor $ F_n $ of mode ($
s_F,i_F $) decreases with increasing values of amplitudes $
\xi_{s_F} $ and $ \xi_{i_F} $. For a certain value of these
amplitudes a minimum value of Fano factor $ F_n $ is reached. In
our case this occurs for approximately 100 incident photons in
both forward-propagating signal and idler fields, as is shown in
Fig. \ref{fig8}.

\begin{figure}    
 \resizebox{0.55\hsize}{!}{\includegraphics{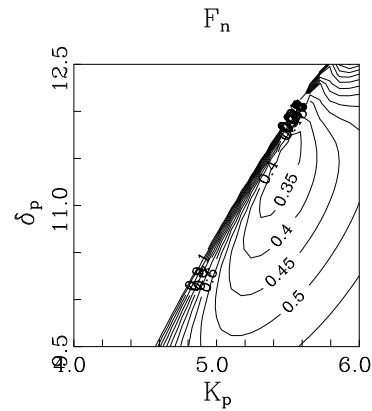}}

 \vspace{2mm}
 \caption{Topological graph of Fano factor $ F_n $
 of mode ($ s_F,i_F $) in dependence on linear coupling constant $ K_p $
 and linear phase mismatch $ \delta_p $ between the pump fields;
 only values of $ F_n $ lower than 1 are plotted;
 $ L=1 $, $ \delta_F = \delta_B = 5 $, $ A_{p_F} = 10 $,
 and values of the other parameters are the same as
 in Fig. 6.}
 \label{fig9}
\end{figure}

If the nonlinear interaction is not phase-matched ($ \delta_{nl} =
\delta_F = \delta_B \neq 0 $) in the waveguide with no photonic
band-gap structure ($ K_s = K_i = K_p = 0 $) sub-Poissonian light
in mode ($ s_F,i_F $) can still be obtained but greater values of
Fano factor $ F_n $ occur. Even values of Fano factor $ F_n $ can
monotonically increase as the value of pump amplitude $ A_{p_F} $
increases in some cases. Then values of parameters of the photonic
band-gap structure ($ K_p \neq 0 $, $ \delta_p \neq 0 $, $ K_s =
K_i = 0 $) can be set in such a way that the original low values
of Fano factor $ F_n $ are restored. If we consider $ \delta_{nl}
= 5 $ mm${}^{-1} $ under the conditions specified in caption to
Fig. \ref{fig6}, the lowest value of Fano factor $ F_n $ is
approximately 0.8. A suitable choice of values of $ K_p $ and $
\delta_p $ provides the original value of Fano factor $ F_n $
being roughly 0.3, as is documented in Fig.\ref{fig9}. According
to our investigations, the region in $ (K_p,\delta_p) $ space for
which the influence of nonlinear phase mismatch $ \delta_{nl} $ is
compensated lies around $ \delta_p \approx 2|K_p| $ and also
greater values of $ K_p $ are needed. As is seen from Fig.
\ref{fig9}, the region where $ \delta_p $ is greater than $ 2|K_p|
$ is not suitable for the generation of sub-Poissonian light.

A suitable choice of the incident phase $ \varphi_{p_F} $ of the
forward-propagating pump mode [$ \varphi_{p_F} = \arg(A_{p_F}) $]
can be used as a final step to reach the lowest possible value of
Fano factor $ F_n $ that is allowed by a given set of values of
waveguide parameters. We have observed a strong dependence of
values of Fano factor $ F_n $ on the value of phase $
\varphi_{p_F} $.

Low values of Fano factor $ F_n $ are usually reached when also
integrated intensity $ W $ of a given mode has low values.

\begin{figure}    
 \resizebox{0.7\hsize}{!}{\includegraphics{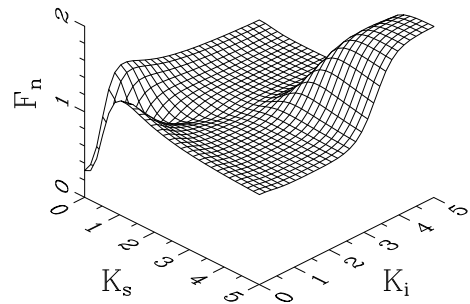}}

 \vspace{2mm}
 \caption{Fano factor $ F_n $
 of mode ($ s_F,i_F $) in dependence on linear coupling constants $ K_s $
 and $ K_i $ between the signal and idler fields;
 $ K_p = 1.4 $, $ A_{p_F} = 10 $,
 and values of the other parameters are the same as
 in Fig. 6.}
 \label{fig10}
\end{figure}

We now consider the influence of photonic band-gap structure in
its full complexity, i.e. also the signal and idler fields are
linearly scattered ($ K_s \neq 0 $, $ K_i \neq 0 $). Under
phase-matched conditions increasing values of $ K_s $ and $ K_i $
destroy sub-Poissonian photon-number statistics in mode ($ s_F,i_F
$) (see Fig. \ref{fig10}).
On the other hand, they support sub-Poissonian-light generation in
modes ($ s_B,i_B $) and ($ s_F,i_B $). Mode ($ s_F,i_B $) can have
values of Fano factor $ F_n $ less than one only for small values
of $ K_s $ and also a nonzero value of $ K_i $ is required. In
general, oscillating regime of behavior of field amplitudes ($
\delta_s > 2|K_s| $, $\delta_i > 2|K_i| $, $ \delta_p > 2|K_p| $)
is more convenient for sub-Poisonian-light generation.

Values of Fano factors $ F_n $ in general depend only on the phase
$ \psi $ [$ \psi = -\arg(K_p) + \arg(K_s) + \arg(K_i) $] that
combines phases of all linear coupling constants. The lowest value
of Fano factor $ F_n $ in mode ($ s_B,i_B $) is reached for $ \psi
= \pi/2 $; $ \psi = 3\pi/2 $ has been found to be optimum for mode
($ s_F,i_B $).

We note that a qualitative similarity can be found in the behavior
of photon-number statistics between the investigated waveguide and
that one composed of two separated nonlinear parts with
co-propagating fields \cite{Mista1999}.

\section{Increase of intensity signal-to-noise ratio}

The nonlinear waveguide can also be used for improving
signal-to-noise ratio of an incident field. This effect can be
explained claiming that the signal part and the noisy part of the
incident field have different amplification coefficients in the
nonlinear process. An average amplification coefficient of a field
in a nonlinear process depends on a statistical distribution of
incident-field phases; there exists one phase for which the
amplification is maximum. If the central phase of the incident
field (corresponding to a coherent signal amplitude) has the
strongest amplification then the noisy part (with a blurred phase
distribution) is less amplified on average and signal-to-noise
ratio increases.

\begin{figure}    
 \resizebox{0.8\hsize}{!}{\includegraphics{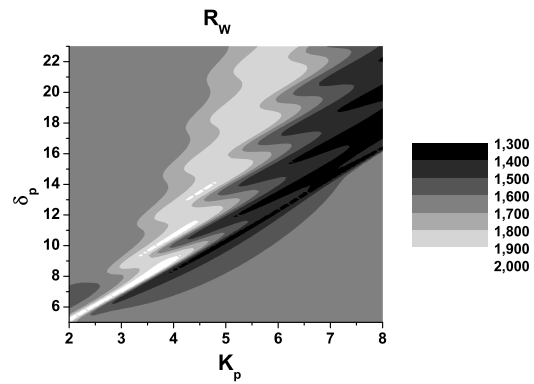}}

 \vspace{2mm}
 \caption{Topological graph of second reduced moment $ R_W $ of integrated
 intensity in mode $ s_F $ as a function of linear coupling constant $ K_p $
 and linear phase mismatch $ \delta_p $ between the pump fields;
 an incident value of $ R_W $ equals 1.75; $ A_{p_F} = 10 $, $ \delta_F = \delta_B =
 5 $, $ \xi_{s_F} = \xi_{i_F} = 10 $, $ n_{{\rm ch},s_F} = 100 $,
 $ n_{{\rm ch},i_F} = 0 $,
 and values of the other parameters are the same as
 in Fig. 6.}
 \label{fig11}
\end{figure}

Similarly as for sub-Poissonian-light generation, the best
conditions for the reduction of noise-to-signal ratio can be found
in a phase-matched waveguide without a photonic band-gap
structure. If nonlinear phase-matching cannot be reached, a
suitable photonic band-gap structure inside the waveguide
compensates for nonlinear phase-mismatch and gives similar
conditions for the reduction of noise-to-signal ratio as those
found in the perfectly phase-matched waveguide. For a given value
of nonlinear phase mismatch $ \delta_{nl} $ ($ \delta_{nl} =
\delta_F = \delta_B $), there are regions in space $ (K_p,
\delta_p) $ where the optimum conditions for the reduction of
noise-to-signal ratio are found. These regions lie around the line
$ \delta_p = 2|K_p| $ and also values of $ K_p $ have to be
greater. Second reduced moment of integrated intensity $ R_W $ [$
R_W = \langle W^2 \rangle_{\cal N} / \langle W \rangle_{\cal N}^2
$] for mode $ s_F $ having incident $ R_W = 1.75 $ (100 incident
signal photons, 100 incident noisy photons) is plotted in Fig.
\ref{fig11} assuming $ \delta_{nl} = 5 $ mm$ {}^{-1} $. Regions
suitable for the reduction of noise-to-signal ratio are visible in
Fig. \ref{fig11}; the lowest achieved value of $ R_W $ is 1.36.

\begin{figure}    
 \resizebox{0.7\hsize}{!}{\includegraphics{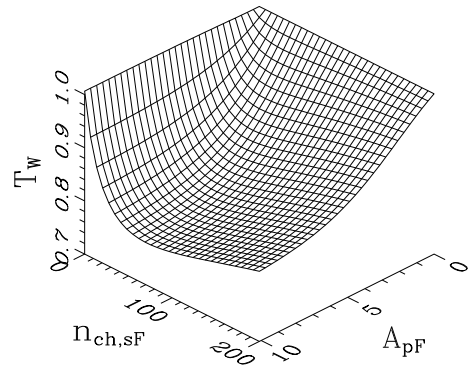}}

 \vspace{2mm}
 \caption{Relative second reduced moment $ T_W $ of integrated
 intensity in mode $ s_F $ as a function of forward-propagating pump
 amplitude $ A_{p_F} $ and number $ n_{{\rm ch},s_F} $ of incident chaotic
 photons in this mode; $ T_W = R_W^{\rm out} / R_W^{\rm in} $ where
 $ R_W^{\rm in} $ ($ R_W^{\rm out} $) characterizes the incident
 (outgoing) field;
 $ K_p = 8 $, $ \delta_p = 20 $, $ \delta_F = \delta_B
 = 5 $, $ \xi_{s_F} = \xi_{i_F} = 10 $, $ n_{{\rm ch},i_F} = 0 $,
 and values of the other parameters are the same as
 in Fig. 6.}
 \label{fig12}
\end{figure}

Capability to reduce noise-to-signal ratio increases with the
increasing forward-propagating pump amplitude $ A_{p_F} $ (see
Fig. \ref{fig12}). This monotonous behavior clearly shows that the
nonlinear interaction is responsible for this effect. The strength
of the effect also depends on the amount of incident noise. As
demonstrated in Fig. \ref{fig12} for mode $ s_F $, an optimum
value of the number of incident noisy photons exists for which the
reduction of noise-to-signal ratio is the best.

\section{Conclusions}

A planar nonlinear photonic band-gap waveguide with optical
parametric process has been analyzed from the point of view of
generation of squeezed light and light with sub-Poissonian
photon-number statistics. It has been shown that squeezed light as
well as sub-Poissonian light can be generated in compound modes
composed of one signal and one idler field; both forward- and
backward-propagating fields can be successfully combined. The
waveguide with a photonic band-gap structure cannot provide better
values of principal squeeze variance and Fano factor in comparison
with the waveguide with no photonic band-gap structure and having
the nonlinear interaction phase-matched. However, if the nonlinear
interaction in a waveguide cannot be phase-matched for some
reason, inclusion of a photonic band-gap structure such as to
compensate for phase mismatch leads to values of principal squeeze
variances and Fano factors that are found assuming perfect
phase-matching. This property makes nonlinear photonic band-gap
waveguides promising as sources of light with nonclassical
properties. Moreover, if a noisy light is incident on the
waveguide its signal-to-noise ratio can be improved as the light
propagates in the waveguide.

\acknowledgments{This work was supported by the COST project OC
P11.003 of the Czech Ministry of Education (M\v{S}MT) being part
of the ESF project COST P11 and by grant LN00A015 of the Czech
Ministry of Education. Support coming from cooperation agreement
between Palack\'{y} University and University La Sapienza in Rome
is acknowledged.}


\end{document}